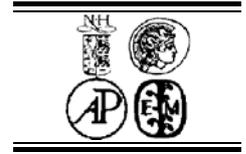

# Performance of a Time-Projection-Chamber with a Large-Area Micro-Pixel-Chamber Readout

Kentaro Miuchi, Kaori Hattori, Shigeto Kabuki, Hidetoshi Kubo, Shunsuke Kurosawa, Hironobu Nishimura, Atsushi Takada, Ken'ichi Tsuchiya, Yoko Okada, Toru Tanimori, Kazuki Ueno

[a]*Kyoto university, Kitashirakawaoiwakecho sakyoku kyoto 606-8502 Japan*

**Abstract**

A micro time-projection-chamber (micro-TPC) with a detection volume of $23\times28\times31$ cm$^3$ was developed, and its fundamental performance was examined. The micro-TPC consists of a micro pixel chamber with a detection area of $31\times31$ cm$^2$ as a two-dimensional imaging device and a gas electron multiplier with an effective area of $23\times28$ cm$^2$ as a pre-gas-multiplier. The micro-TPC was operated at a gas gain of 50,000, and energy resolutions and spatial resolutions were measured.

## 1. Introduction

Large volume time projection chambers (TPC) have been developed for various applications[1,2,3]. Micro patterned gaseous detectors, such as a gas electron multiplier (GEM) [4], a micromegas[5] and a micro pixel chamber[6], are thought to improve the spatial resolutions of TPCs. Previously, we have developed a micro-time-projection-chamber (micro-TPC) with a detection volume of $10\times10\times8$ cm$^3$ [7] based on a micro pixel chamber (μ-PIC) with a detection area of $10\times10$ cm$^2$. This micro-TPC proved several conceptual ideas for such applications as an electron-tracking Compton camera [8], a dark matter detector [9], and a neutron-imaging detector [10]. In order to achieve the physical goals of these applications, we now have developed a larger volume micro-TPC based on a large-size μ-PIC with a detection area of $31\times31$ cm$^2$ [11] and a large-size gas electron multiplier (GEM) with an effective area of $23\times28$ cm$^2$. In this paper, the fundamental performance of the large-volume micro-TPC with a detection volume of $23\times28\times31$ cm$^3$ is described.

## 2. Large-volume micro-TPC

### 2.1. Large-area μ-PIC and large-area GEM

A second generation large-area μ-PIC with a detection area of $31\times31$ cm$^2$ was used for this micro-TPC (TOSHIBA, S/N 20060222-3). The production yield was increase to 90% as the result of an improvement of the plating technology. The fractions of dead and bad anode electrodes were decreased to less than 0.1%. The operation gas gain of the μ-PIC was about 5000. The production yield, the fraction of the dead and bad electrodes, and the operation gas gain of the first production μ-PICs were 50%, 5%, and 3500, respectively [11].

A gas electron multiplier (GEM) with an effective area of $23\times28$ cm$^2$ was developed as a pre-multiplication device of the micro-TPC. The GEM was made by a Japanese company, Scienergy Co. Ltd. The size was currently restricted by the working size for the fabrication. Copper electrodes were formed on both sides of the polyimide insulator. The holes were 50 μm in diameter and were place with a pitch of 140 μm. We supplied a voltage of 470 V between the top electrode and the bottom electrode of the GEM in a dry nitrogen gas atmosphere for 30 minutes before use. This aging process prevented the GEM from serious discharges.

A dedicated readout system for the micro-TPC was developed [12]. TPC signals read by 768 anode-strips and 768 cathode-strips were digitized in amplifier shaper discriminator (ASD) chips [13], synchronized in a position encoding system at 100 MHz clock, and recorded by a VME-bus memory-board. Each digital signal or "hit" was a set of ($X_{min}$, $X_{max}$, $Y_{min}$, $Y_{max}$, T), where $X_{min}$ and $X_{max}$ were the minimum and maximum positions of the anode strips, which detected the TPC signals, $Y_{min}$, $Y_{max}$ were those of the cathode strips, and T was the clock counter. The position and a size of an electron cloud (400 μm/digit), and the elapsed time from the trigger of each hit (10 ns/digit) were recorded, so that a track of a charged particle was recorded as successive hits. TPC signals read by 768 cathode-strips were amplified in the ASD chips and were summed into eight channels (each channel had 96 strips), and their waveforms were recorded by a 100 MHz flash-ADC. The energy deposition of a charged particle was thus known from the detected waveforms.

*2.2. Large-volume micro-TPC*

A schematic drawing of the micro-TPC is shown in Fig. 1. The micro-TPC had a drift length of 31 cm, which made a detection volume of $23\times28\times31$ cm$^3$. The gas volume was set in an aluminum vessel filled with a normal-pressure argon-ethane gas mixture (9:1). A drift voltage of -7.5 kV was supplied to the drift plane ($V_{DRIFT}$=7.5kV), which made a drift electric field of 0.23 kV/cm. The voltage of the top plane of the GEM was set at -660 V ($V_{GEMT}$=660 V), which was reconciled with the voltage of the shaping pattern at the corresponding drift length. The voltage of the bottom plane of the GEM was determined by the GEM voltage ($V_{GEMB} = V_{GEMT} - \Delta V_{GEM}$). The distance between the GEM and the μ-PIC was 5 mm.

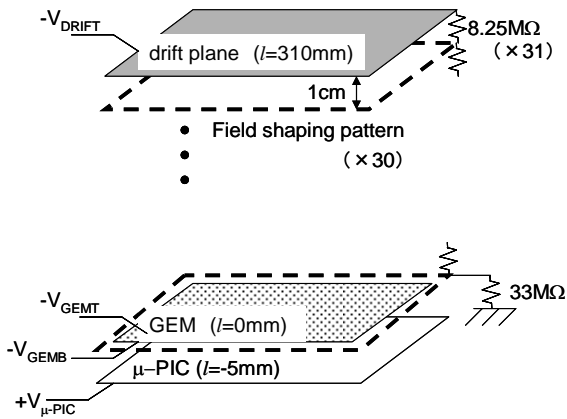

Fig. 1. Schematic drawing of the micro-TPC system. The sizes of the drift plane, field shaping pattern, GEM, and the μ-PIC are

### 3. Performance

*3.1. Gas gains*

The gas gains were measured by irradiating the micro-TPC with X-rays from a radioactive source of $^{109}$Cd. We calculated the collected charge from the flash-ADC data, while the corresponding event position was known from the digital hit data. In order to evaluate any inhomogeneity of the gas gains of the whole detection area, we divided the detection area into $6\times8$ parts, and calculated the gas gain for each part. The ratio of the maximum gas gain to the minimum one was 2.2. The gas gains at the gain-maximum area, measured with several voltages supplied to the anode electrodes of the μ-PIC ($V_{\mu-PIC}$) and several voltages supplied to the GEM ($\Delta V_{GEM} = V_{GEMT} - V_{GEMB}$), are shown in Fig. 2. The internal gas gain of the μ-PIC operated without the GEM is shown for a comparison [11]. The internal gain without the GEM was about 5000. When we increased $\Delta V_{GEM}$, higher total gas gains were observed. A gas gain of 50,000 was achieved with a μ-PIC voltage of 520 V and a GEM voltage of 260 V. This was the required gas gain for tracking the minimum ionizing particles (MIPs) with a 400μm-pitch detector. We therefore operated the micro-TPC with these voltages in the following measurements.

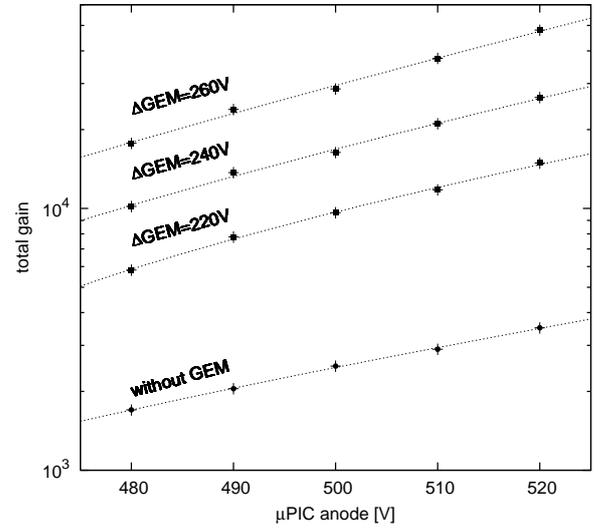

Fig.2. Gas gains as a function of the voltage supplied to the anode electrodes of the μ-PIC. The gas gains measured with GEM voltages of 220V, 240V, and 260V are shown. The gas gains measured without the GEM are also shown for a comparison. Best fit exponential functions are also shown fro eye-guiding.

*3.2. Energy resolutions*

A typical spectrum obtained by the irradiation of X-rays from a radioactive source of $^{133}$Ba is shown in Fig. 3. The whole volume of the micro-TPC was irradiated. The flash-

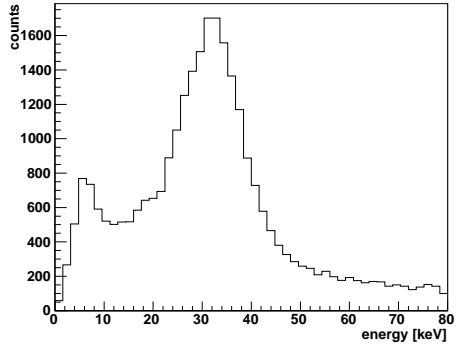

Fig.3. Energy spectrum taken by the irradiation γ-rays form a $^{133}$Ba radioactive source.

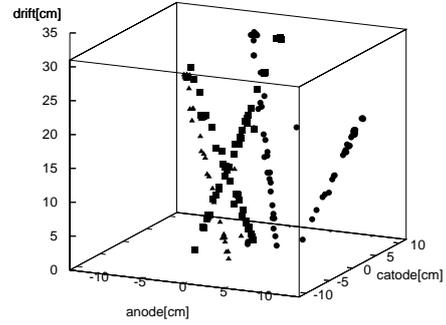

Fig. 5. Muon tracks taken by the micro-TPC. Each mark shows one "digital hit".

ADC data were summed with an energy correction corresponding to the event position to compensate the inhomogeneity of the gas gains. We can see the peak of direct X-rays at 31 keV and the peak of copper fluorescent X-rays from the GEM and the μ-PIC at 8.0 keV. The energy resolutions were measured for the energy range between 8 keV and 60 keV with radioactive sources of $^{109}$Cd, $^{133}$Ba, and $^{241}$Am. The measured energy resolutions are shown in Fig. 4. The energy resolution was 60% FWHM for 60 keV γ-rays. An improvement of the plating technology is expected to bring the energy resolution at worst to 30%, which we achieved with a 10 cm micro-TPC.

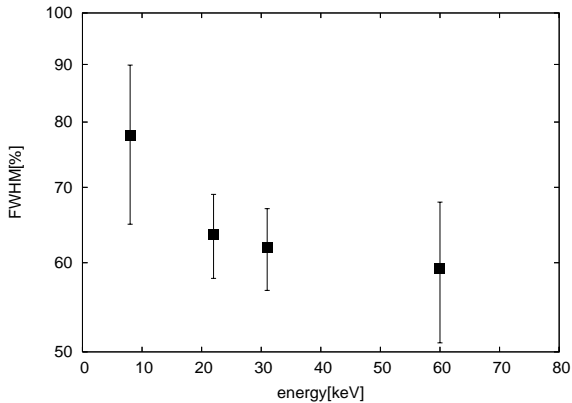

Fig. 4. Energy resolution dependence on the energy.

### 3.3. Position resolution

We measured the three-dimensional position resolution of the micro-TPC by cosmic-ray muons. We placed two plastic scintillators (30×30×1 cm$^3$) above and below the micro-TPC. The micro-TPC was triggered by a coincidence signal of these two scintillators.

First, we measured the drift velocity. The maximum time was 7.3 μs, which indicates a drift velocity of 4.2 cm/μs, taking account of the corresponding drift length of 31 cm. Typical three-dimensional muon tracks reconstructed with the measured drift velocity are shown in Fig. 5. We then fit the muon tracks with straight lines, and the residual was calculated for each hit point. Here, the residual was defined as the distance between a hit point and the fitted straight line. The hit points were divided into 31 groups by their drift length and we plotted the residual distribution for each group. The residual distribution for the hits of *l*=5.5±0.5 cm is plotted in Fig. 6. Each residual distribution was then fitted with a two-dimensional Gaussian. given by (1),

$$f(r) = A \frac{\sqrt{2\pi}}{\sigma^2} r \exp\left(-\frac{r^2}{2\sigma^2}\right) dr \qquad (1)$$

where A is the normalizing factor, *r* is the residual and σ$^2$ is the dispersion of the Gaussian. We obtained σ=1.0 mm for *l*=5.5±0.5 cm. σ was expected to be the root sum square of the detector intrinsic term, σ$_{detector}$, and the diffusion term, σ$_{diffusion}$, as shown in equation (2).

$$\sigma^2(l) = \sigma_{detector}^2 + \sigma_{diffusion}^2 = \sigma_{detector}^2 + \left(D\sqrt{l}\right)^2 \qquad (2)$$

Therefore σ was plotted as a function of *l* (Fig. 7), and fitted with equation (2). As a result of the fitting, σ$_{detector}$=0.51 mm and D=0.37 mm·cm$^{-1/2}$ were obtained. The detector intrinsic term was worse than the value of 0.37 mm that we measured with a 10-cm micro-TPC[11]. The distortion of the spatial resolution could be due to the inhomogeneity of the anode electrodes. The diffusion term is reasonable when we consider that the transverse diffusion is 0.5 mm·cm$^{-1/2}$ and the center position of the digital hit positions was calculated.

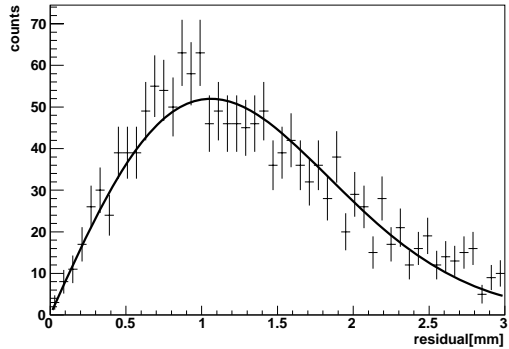

Fig. 6. Residual distribution of the muon tracks for $l=5.5\pm0.5$ cm.

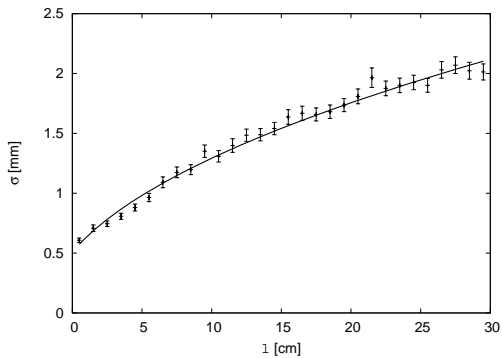

Fig. 7 $\sigma$ dependence on the drift length.

## 4. Prospects

Development issues for applications were found from the above measurements. For the electron-tracking Compton camera, we need to widen the energy range up to about 200 keV. High-energy peaks will be observed by an improvement of the energy resolution. The plating technology, which is still being improved, is expected to produce more homogeneous anode electrodes in the next production. This will improve the homogeneity of the gas gain, and thus the energy resolution will improve. The angular resolutions and the detection efficiency of the Compton camera are expected to be improved. For a dark matter search experiment, stable operation with a lower pressure gas of a few tens of torr should be confirmed. A better spatial resolution is expected with $CF_4$ gas, which will help to improve the sensitivity to dark matters. Operation with $^3$He gas will show potential for the time-resolved neutron imaging detector.

## 5. Conclusions

A micro-TPC with a detection volume of $23\times28\times31$ cm$^3$ was developed. The micro-TPC was operated at a gas gain of 50,000 and an energy resolution of 60% FWHM at 60 keV was measured. We detected tracks of MIPs with the micro-TPC and a three-dimensional position resolution of 0.51 mm was measured.

## Acknowledgement

This work was supported by a Grant-in-Aid in Scientific Research of the Japan Ministry of Education, Culture, Science, Sports and Technology, SENTAN of Japan Science and Technology Agency, and a Grant-in-Aid for the 21st Century COE "Center for Diversity and Universality in Physics". This work is partially supported by a research-aid program of Toray Science Foundation.